\documentstyle[12pt]{article}
\newcommand{\bb}{\begin{eqnarray}}
\newcommand{\ee}{\end{eqnarray}}
\begin{document}
\title{Thermodynamics of 
charged anti-de Sitter black holes in canonical ensemble}
\author{P. Mitra\thanks{On leave from Saha Institute of Nuclear Physics,
Calcutta 700 064;
e-mail address Parthasarathi.Mitra@cern.ch, mitra@tnp.saha.ernet.in}\\
Theory Division, CERN, CH-1211 Geneva 23}
\date{gr-qc/9903078}
\maketitle
\begin{abstract}
As in the grand canonical treatment of  Reissner - Nordstr\"{o}m black holes
in anti-de Sitter spacetime, the canonical ensemble formulation also
shows that non-extremal black holes tend to have lower action than
extremal ones. However, some small non-extremal black holes
have higher action, leading to the possibility of transitions
between non-extremal and extremal black holes. 
\end{abstract}

\bigskip
Black hole thermodynamics continues to be an interesting field of study.
A special area of recent interest is that of {\it extremal} black holes.
While the entropy of ordinary ({\it non-extremal}) black holes has been
taken to be a quarter of the horizon area for a long time, recently there
has been some confusion in the case of {\it extremal} black holes. The semiclassical
derivations of the entropy formula for non-extremal black holes
do not directly apply to extremal black holes, and because of
the difference in topology of euclidean extremal and
non-extremal black holes, one cannot rely on extrapolation. In fact,
euclidean studies indicate that extremal black holes should have zero entropy
\cite{HHR} even though the horizon area is nonzero.
On the other hand, practical-minded people have tended to expect that extremal
black holes should satisfy the area law just like non-extremal black holes
from which they differ ever so slightly.
One way of accommodating their point of view is 
to argue that there may be different ways of looking at extremal black holes.
Usually, when one quantizes a classical theory, one
tries to preserve the classical topology. In this spirit, one seeks
to have a quantum theory of extremal black holes based exclusively on
extremal topologies. As an alternative, one can have a quantization where
a summation is carried out over topologies.  Then, in the consideration of the
functional integral, classical configurations corresponding to both
topologies must be included \cite{GM}. The extremality condition can  subsequently be
imposed on the averages that result from the functional integration. It has been
customary, following \cite{york}, to use a grand canonical
ensemble. Here the temperature and the potential for the charges
are supposed to be specified as inputs, and the average mass $M$ and charges
$Q$ of the black hole are outputs. So the actual definition of extremality
that is involved here for a Reissner - Nordstr\"{o}m black hole with one kind
of charge is $Q=M$.  This may be described  as {\it extremalization after
quantization}, as opposed to the usual approach of {\it quantization after
extremalization}.  It was shown in \cite{GM} that
extremalization after quantization does lead to an entropy equal to a
quarter of the area.
But does the approach of quantization after extremalization
lead to zero entropy?
Even that is not quite true \cite{com} for the extremal  
Reissner - Nordstr\"{o}m black hole: the reason is that the
semiclassical approximation fails because the action does not have a stable
minimum there. However, if an
asymptotically anti-de Sitter version of the extremal Reissner -
Nordstr\"{o}m black hole is considered,  a stable
minimum does occur \cite{eads}. Consequently, there is a sensible
semiclassical approximation, and as expected in \cite{HHR}, the entropy
vanishes if quantization is carried out after extremalization.  On the other hand, if
quantization is carried out first, the entropy is once again given by a
quarter of the area. Thus, in an asymptotically anti-de Sitter spacetime, two
different kinds of extremal charged black holes exist: those obtained by
quantization after extremalization, and those obtained by reversing the order
of these operations. This  has been confirmed in a hamiltonian
framework \cite{louko}.

The conclusions of \cite{eads} were reached in a grand canonical treatment. We shall
go on to do a similar analysis here in the canonical ensemble. There are two
motivations for this. First, it is known that the different thermodynamical ensembles
are not exactly equivalent and may not lead to the same conclusions as they
correspond to different physical situations. Secondly, the canonical analysis 
in this system allows the consideration of the kind of transition discussed
in \cite{HP} between neutral black holes in anti-de Sitter spacetime and pure
anti-de Sitter spacetime. If charged black holes are envisaged, they can be
imagined to decay into other charged configurations \cite{cvetic}. Extremal black holes
defined by quantization after extremalization are well suited for this r\^{o}le
because, as in the case of pure anti-de Sitter spacetime, the euclidean time
coordinate here can be given an arbitrary periodicity. The same cannot
be said about extremal black holes defined by quantization before 
extremalization, of course.

The   Reissner   -   Nordstr\"{o}m   black  hole   solution    of
Einstein's  equations  in free space with a negative cosmological
constant $\Lambda=-{3\over l^2}$ is given by
\bb
ds^2=-hdt^2+h^{-1}dr^2+r^2d\Omega^2, ~A={Q\over r}dt,
\ee
with
\bb
h= 1-{r_+\over r}-{r_+^3\over l^2r}-{Q^2\over r_+r}
+{Q^2\over r^2}+{r^2\over l^2}.
\ee
The asymptotic form of this spacetime is anti-de Sitter.
There is an outer horizon  located at $r=r_+$.
The mass of the black hole is given by
\bb
M={1\over 2}(r_++{r_+^3\over l^2}+{Q^2\over r_+}).\label{M}
\ee
It satisfies the laws of black hole thermodynamics with a temperature
\bb
T_H={1-{Q^2\over r_+^2} +{3r_+^2\over l^2}\over 4\pi r_+}\label{T}
\ee
and a potential
\bb
\phi={Q\over r_+}.
\ee
In general $r_+,Q$ are independent, but in the extremal case they get
related:
\bb
1-{Q^2\over r_+^2} +{3r_+^2\over l^2}=0.\label{ex}
\ee

The usual action for the euclidean version of the anti-de Sitter
Reissner - Nordstr\"{o}m
black hole on a four dimensional manifold ${\cal M}$ with a boundary
is given by
\bb
I&=&-{1\over  16\pi}\int_{\cal M} d^4x\sqrt g(R-2\Lambda)
+{1\over 8\pi}\int_{\partial
{\cal M}} d^3x\sqrt\gamma (K-K_0)\nonumber\\
&+&{1\over 16\pi}\int_{\cal M}
d^4x\sqrt g F_{\mu\nu} F^{\mu\nu}.\label{action}
\ee
Here  $\gamma$ is the induced metric on the boundary $\partial {\cal M}$
and $K$ the extrinsic curvature of the boundary. $K_0$ is to be
chosen to make the action finite.
We shall  study the action for off-shell configurations near the
black hole solution. For simplicity, only
a class of spherically symmetric
metrics \cite{york} is considered on ${\cal M}$:
\bb
ds^2=b^2d\tau^2+\alpha^2dr^2+r^2d\Omega^2,
\ee
with  the variable $r$ ranging between $r_+$ (the horizon) and $r_B$ (the
boundary), and $b, \alpha$ functions of $r$  only.  There  are
boundary conditions as usual \cite{york,GM,eads}:
\bb
b(r_+)=0,~2\pi b(r_B)=\beta.
\ee
This corresponds to the convention of fixing the range of integration
of the euclidean time $\tau$ to be $2\pi$.
$\beta$  is the inverse temperature at the boundary of radius $r_B$.
There is another boundary condition involving $b'(r_+)$:
It reflects the extremal/non-extremal  nature  of  the  black  hole  and  is
therefore   different   for  the  two cases:
\bb
{b'(r_+)\over\alpha(r_+)}&=&1{\rm ~in~non-extremal~case},\nonumber\\
&{\rm and}& 0 {\rm ~in~extremal~case}.
\ee

In the spherically symmetric situation, the  vector
potential was taken to be zero (a radial component may be gauged away)
and the scalar potential required to  satisfy
the boundary conditions
\bb
A_\tau(r_+)=0, ~A_\tau(r_B)={\beta\phi\over 2\pi i}.
\ee
The boundary condition at $r_B$ fixes the potential there and thus corresponds
to the choice of the grand canonical ensemble. To go to the canonical
ensemble, as we propose to do here, the electric field (or enclosed
charge) rather than the potential
has to be fixed at $r_B$. This involves altering the action \cite{york} with
a boundary term so that the appropriate variational principle yields the
equations of motion.
The action (\ref{action}) with  the above metric  
takes the form:
\bb
I&=&{1\over      2}\int^{2\pi}_0      d\tau\int^{r_B}_{r_+}      dr
\bigg(-{2rb'\over\alpha}  -{b\over\alpha}  -\alpha  b   +\Lambda\alpha
br^2\bigg)         -{1\over         2}\int^{2\pi}_0              d\tau
\bigg[{(br^2)'\over\alpha}\bigg]_{r=r_+}\nonumber\\ &+& I_0
+ {1\over      2}\int^{2\pi}_0      d\tau\int^{r_B}_{r_+}      dr
{r^2\over\alpha b}{A'}_\tau^2.
\ee
$I_0$ is the contribution of the $K_0$ term in the action and has to be chosen
so as to make the action finite in the limit of large $r_B$ with $\beta$
appropriately scaled.
The conversion of this action to the form appropriate to the canonical ensemble simply requires
the addition of a piece
$-2\pi{r^2\over\alpha b}A'_\tau A_\tau|_{r_B}$.
Variation of this modified action with the functions $ b(r), \alpha(r)$ and $A_\tau(r)$ under
boundary  conditions  appropriate to the new situation leads  to  reduced versions of
the  Einstein  -  Maxwell
equations. The solution of a subset of these equations,
namely the Gauss law and the hamiltonian constraint, is given by
\cite{york,RNads}
\bb
{1\over\alpha}=\bigg(1-{r_+\over r}-{r_+^3\over l^2r}-{q^2\over r_+r}
+{q^2\over r^2}+{r^2\over l^2}\bigg)^{1/2},
\quad A'_\tau=-{iqb\alpha\over r^2},
\ee
with  $r_+$ and  $q$ arbitrary at this stage. The value of $q$ has to be fixed to
define the canonical ensemble and the potential is not to be treated as being specified at
the boundary in this ensemble.
The above action with the boundary term added may be expressed in terms of $r_+$ and then
has to be extremized with respect to $r_+$ as in \cite{york}.
The value of the action is
\bb
I&=& -\beta r_B\sqrt{1-{r_+\over r_B}-{r_+^3\over l^2r_B}-
{q^2\over r_+r_B}
+{q^2\over r_B^2}+{r_B^2\over l^2}} +I_0-\pi r_+^2
{\rm ~(non-ext~bc)},\nonumber\\
&{\rm and}& -\beta r_B\sqrt{1-{r_+\over r_B}-{r_+^3\over l^2r_B}-
{q^2\over r_+r_B}
+{q^2\over r_B^2}+{r_B^2\over l^2}}+I_0
{\rm ~(ext~bc)}.\label{I}
\ee
The first line is analogous to \cite{york,RNads},
where the non-extremal condition was used
in connection with a semiclassically quantized  non-extremal  black
hole. The second line is similar to  the consequence of the
extremal condition
used in connection with  a  semiclassically
quantized   extremal   black   hole
\cite{com,GM,eads}. The only difference of (\ref{I}) with the corresponding 
equation in \cite{eads} is the absence of a $q\beta\phi$ term which got
removed when a surface term was added to alter the nature of boundary data
from the grand canonical type to the canonical type. 

The above ``reduced action'' has to be extremized with respect to $r_+$
in order to impose the equations of motion ignored so far.

We shall consider several cases:

\bigskip

{\bf 1)} The extremization of (\ref{I}) with respect to $r_+$ in the {\it
non-extremal case} yields the relation
\bb
{\beta \over\sqrt{
1-{r_+\over r_B}-{r_+^3\over l^2r_B}-{q^2\over r_+r_B}
+{q^2\over r_B^2}+{r_B^2\over l^2}
}}= {4\pi r_+\over (1-{q^2\over r_+^2}+{3r_+^2\over l^2})}.\label{r_+}
\ee
This relation can be considered
to fix $r_+$ in terms of the specified value of  $\beta$;
conversely, it also shows the expected form of $\beta$ as a function of
$r_+$. It may be noted that the value of $r_+$ with a
given value of the left hand side (${\beta l\over r_B}$ for large $r_B$) is unique
only for large $|q| (>{l\over 6})$. For smaller $|q|$, the equation for $r_+$
has three positive solutions for certain values of $\beta$.
The second derivative of the action with respect to $r_+$ does not in
general have a definite sign, being equal, for large $r_B$, to
\bb
{2\pi ({3r_+^2\over l^2}+{3q^2\over r_+^2}-1)\over 1-{q^2\over l^2}+
{3r_+^2\over l^2}}.
\ee
The numerator can be guaranteed to be positive for large $|q| (>{l\over 6})$ only.
For smaller $|q|$,
some non-extremal black hole solutions may not be minima of the
classical action.
The entropy corresponding to the saturation of the partition function
by an extremum can be confirmed to be
\bb
S&=&\beta^2{d (I/\beta)\over d\beta}\nonumber\\ &=
&\beta^2 {d\over d\beta}\bigg(-r_B\sqrt{1-{r_+\over r_B}-{r_+^3\over l^2r_B}-
{q^2\over r_+r_B}
+{q^2\over r_B^2}+{r_B^2\over l^2}}+{I_0\over\beta}-{\pi r_+^2\over\beta}\bigg)
\nonumber\\ &=&\beta^2({\pi r_+^2\over\beta^2}-{I_0\over\beta^2}) +\beta{dI_0\over d\beta}
\nonumber\\ &-&\beta^2{dr_+\over d\beta}{d\over dr_+}
\bigg(-r_B\sqrt{1-{r_+\over r_B}-{r_+^3\over l^2r_B}- {q^2\over r_+r_B}
+{q^2\over r_B^2}+{r_B^2\over l^2}}
-{\pi r_+^2\over\beta}\bigg)\nonumber\\ &=&\pi r_+^2,
\ee
where the $I_0$ terms cancel because of linear 
homogeneity in $\beta$ and the $r_+$ derivative
vanishes because it defines the extremum of the action.
Thus the area formula is valid here.

\bigskip

{\bf 2)} The extremization of (\ref{I}) can be done for the 
{\it extremal} condition, where,
however, the action is homogeneous in $\beta$, which disappears from
the relation fixing $r_+$. This is not surprising: in the extremal case $q,r_+$ are
known to be related to each other by (\ref{ex}), and the temperature
is arbitrary as there is no conical singularity \cite{HHR}.
The second derivative of the action with respect to
$r_+$ is proportional to ${3r_+^2\over l^2}+{q^2\over r_+^2}$, which is
positive definite, so the extremal black hole solutions are strict minima of
the classical action. 
%This behaviour contrasts with the asymptotically flat case
%\cite{com}. 
The grand canonical calculation led to a similar result \cite{eads}
for large $r_B$ and finite $l$. The canonical result holds for all $r_B$
and persists in the limit $l\to\infty$.
The entropy corresponding to the saturation of the action by this
minimum is zero. This follows from the fact \cite{HHR} that 
the action continues to be proportional to $\beta$ after the extremizing
value of $r_+$ is plugged in. Hence,
\bb
S=\beta^2{d(I/\beta)\over d\beta}=0.
\ee
Thus this case refers to an extremal black hole of {\it arbitrary} temperature
and zero entropy.

\bigskip

{\bf 3)} The previous case refers to the quantized extremal black hole.
As in \cite{GM,eads}, there is a possibility of quantizing the black
hole {\it before} extremizing it, {\it i.e.,} the
two topologies may be summed over in the functional integral 
and the extremality condition imposed afterwards on the averaged quantities.
The partition function is of the form
\bb
\sum_{\rm topologies}\int d\mu(r_+) e^{-I(r_+,{\rm topology})},
\ee
with   $I$   given    by    (\ref{I})    as    appropriate    for
non-extremal/extremal topology.
The  semiclassical approximation involves
replacing the double integral by the
maximum value of the integrand, {\it i.e.,} by the
exponential of the negative of the minimum  $I$.
One has to consider the variation of $I$ as $r_+$ varies in both topologies.
It  is  clear from (\ref{I})
that the non-extremal action can be made lower than the extremal one
because of the extra term $-\pi r_+^2$.
Consequently, the partition function is  to  be  approximated  by
$e^{-I_{min}}$,  where $I_{min}$ is the classical
action for the {\it non-extremal} case,
{\it minimized} with respect to $r_+$.
As in the non-extremal case, this leads to an entropy equal to a quarter
of the horizon area. Extremality is imposed eventually through the condition
(\ref{ex}) on $r_+$. The two requirements on $r_+$ become consistent only
in the limit $\beta\to\infty$. Thus this case refers to an extremal black hole
of zero temperature and entropy equal to a quarter of the area of the horizon.
It corresponds to a different way of formulating the extremal black hole
from the previous case. This is the approach of {\it quantization before extremalization}
whereas the earlier one was the pure extremal approach: {\it extremalization before
quantization}. 

\bigskip

{\bf 4)} The previous case involved a comparison of non-extremal and extremal
configurations geared towards the definition of extremal black holes. One can make a
more direct comparison of the actions corresponding to the first two cases.
If we denote by $r_+$ the radius of the horizon of the non-extremal black hole
as in case 1 above and refer to the corresponding quantity for the extremal black
hole as in case 2 above by 
\bb
r_0\equiv l\sqrt{\sqrt{1+{12q^2\over l^2}}-1\over 6}, 
\ee
we see that
\bb
I_{\rm non-ex}-I_{\rm ex} &=&
-\beta r_B\sqrt{1-{r_+\over r_B}-{r_+^3\over l^2r_B}-
{q^2\over r_+r_B}
+{q^2\over r_B^2}+{r_B^2\over l^2}}-\pi r_+^2\nonumber\\ &&
+\beta r_B\sqrt{1-{r_0\over r_B}-{r_0^3\over l^2r_B}-
{q^2\over r_0r_B}
+{q^2\over r_B^2}+{r_B^2\over l^2}}.
\ee
The $I_0$ terms, which depend only on $\beta$ and $r_B$ have been cancelled
out here. For large $r_B$, by making use of (\ref{r_+}), we get
\bb
I_{\rm non-ex}-I_{\rm ex}=
{4\pi r_+\over 1-{q^2\over r_+^2}+{3r_+^2\over l^2}}\bigg( {r_+\over 2}
+{q^2\over 2r_+}+{r_+^3\over 2l^2}- {r_0\over 2}
-{q^2\over 2r_0}-{r_0^3\over 2l^2}\bigg) -\pi r_+^2,
\ee
which can be recognized to be the difference of the free energies of
the two black holes if one remembers the expression (\ref{M}) for the
mass of a black hole and the fact that the extremal black hole being
considered here is of the pure type with zero entropy.
On simplification,
\bb
I_{\rm non-ex}-I_{\rm ex}=
{\pi(r_+-r_0)\over 1-{q^2\over r_+^2}+{3r_+^2\over l^2}}\bigg(
r_+-3r_0-{r_+^3+r_+^2r_0+r_+r_0^2+9r_0^3\over l^2}\bigg).
\ee
For a black hole with positive temperature, $r_+>r_0$, so that
the sign of this difference depends on the last factor involving
cubics in $r_+$ and $r_0$. For large enough $r_0$, this expression is
negative for all allowed $r_+$, which means that all non-extremal black holes
are stable against decay into the extremal black hole. However, for
small $r_0$, {\it i.e.,} for small charge, there exists a range of values
of $r_+$ for which the factor is positive, corresponding to the
occurrence of non-extremal black holes capable of decaying to extremal
black holes. The transition to this small charge behaviour from the
large charge behaviour occurs at the positive real root of
\bb
516r_0^6+392r_0^4l^2+77r_0^2l^4-l^6=0,
\ee
which is approximately given by
\bb
r_0\approx 0.1105 l,
\ee
corresponding to a charge of
\bb
|q_0|\approx 0.1125 l<{l\over 6}.
\ee

In conclusion, we have found that the results of the grand canonical
calculations of \cite{eads} are mostly reproduced in the canonical ensemble:
extremal black holes in an asymptotically anti-de Sitter spacetime
can be defined in two ways, one having  zero entropy and arbitrary temperature
and the other having zero temperature and finite entropy. However, while a non-extremal
black hole in such spacetimes usually has lower free energy, in some special
cases of small size it has higher free energy and can decay into extremal black holes.

\section*{Acknowledgment}

This work was completed in the Theory Division of CERN, whose support
is gratefully acknowledged.

\end{document}